\documentclass{ws-ijmpd}
\usepackage[super,compress]{cite}
\usepackage{amsmath,amssymb,latexsym}
\usepackage{graphicx}

\newcommand\rf[1]{(\ref{eq:#1})}
\newcommand\lab[1]{\label{eq:#1}}
\newcommand\nonu{\nonumber}
\newcommand\br{\begin{eqnarray}}
\newcommand\er{\end{eqnarray}}
\newcommand\be{\begin{equation}}
\newcommand\ee{\end{equation}}

\newcommand\foot[1]{\footnotemark\footnotetext{#1}}
\newcommand\lb{\lbrack}
\newcommand\rb{\rbrack}

\newcommand\llb{\left\lbrack}
\newcommand\rrb{\right\rbrack}

\renewcommand\({\left(}
\renewcommand\){\right)}

\newcommand\bc{\begin{center}}
\newcommand\ec{\end{center}}

\newcommand\partder[2]{\frac{{\partial {#1}}}{{\partial {#2}}}}

\renewcommand\a{\alpha}

\renewcommand\d{\delta}

\newcommand\eps{\epsilon}
\newcommand\vareps{\varepsilon}

\newcommand\G{\Gamma}

\newcommand\h{\frac{1}{2}}
\renewcommand\k{\kappa}
\renewcommand\l{\lambda}
\renewcommand\L{\Lambda}
\newcommand\m{\mu}
\newcommand\n{\nu}

\newcommand\vp{\varphi}
\renewcommand\P{\Phi}
\newcommand\pa{\partial}

\newcommand\pr{\prime}

\newcommand\s{\sigma}

\renewcommand\t{\tau}

\newcommand\cA{{\mathcal A}}
\newcommand\cB{{\mathcal B}}

\newcommand\cH{{\mathcal H}}

\newcommand{\ct}[1]{\cite{#1}}
\newcommand{\bib}[1]{\bibitem{#1}}

\newcommand\PRL[3]{\textsl{Phys. Rev. Lett.} \textbf{#1} (#2) #3}

\newcommand\PRD[3]{\textsl{Phys. Rev.} \textbf{D#1} (#2) #3}

\newcommand\CQG[3]{\textsl{Class. Quantum Grav.} \textbf{#1} (#2) #3}

\newcommand\IJMPA[3]{\textsl{Int. J. Mod. Phys.} \textbf{A#1} (#2) #3}

\newcommand\MPLA[3]{\textsl{Mod. Phys. Lett.} \textbf{A#1} (#2) #3}

\newcommand\udot{\stackrel{.}{u}}

\newcommand\Hdot{\stackrel{.}{H}}

\newcommand\Adot{\stackrel{.}{A}}
\newcommand\Bdot{\stackrel{.}{B}}

\begin{document}

\markboth{Eduardo Guendelman, Emil Nissimov, Svetlana Pacheva}
{Gravity-Assisted Emergent Higgs Mechanism in the Post-Inflationary Epoch}

%
\catchline{}{}{}{}{}
%

\title{\textbf{GRAVITY-ASSISTED EMERGENT HIGGS MECHANISM IN THE 
POST-INFLATIONARY EPOCH}
\foot{Honorable Mention in the Gravity Research Foundation Essay Competition 2016}
}

\author{Eduardo Guendelman}

\address{Department of Physics, Ben-Gurion University of the Negev,
Beer-Sheba 84105, Israel \\
guendel@bgu.ac.il}

\author{Emil Nissimov, Svetlana Pacheva}
\address{Institute for Nuclear Research and Nuclear Energy,\\
Bulgarian Academy of Sciences, Sofia 1784, Bulgaria\\
nissimov@inrne.bas.bg , svetlana@inrne.bas.bg}

\date{March 20, 2016}
\maketitle

\maketitle

\begin{history}
\received{}
\revised{}
\end{history}

\begin{abstract}
We consider a non-standard model of gravity coupled to a neutral scalar
``inflaton'' as well as to $SU(2)\times U(1)$ iso-doublet scalar with positive mass
squared and without self-interaction, and to  $SU(2)\times U(1)$ gauge fields. 
The principal new ingredient is employing two alternative non-Riemannian 
space-time volume-forms (covariant 
integration measure densitities) independent of the metric. The latter have
a remarkable impact -- although not introducing any additional propagating
degrees of freedom, their dynamics triggers a series of important features:
appearance of infinitely large flat regions of the effective ``inflaton'' 
potential as well as dynamical generation of Higgs-like spontaneous symmetry breaking 
effective potential for the $SU(2)\times U(1)$ iso-doublet scalar.
\end{abstract}


\ccode{PACS numbers:04.50.Kd, 
11.30.Qc, 
}

\section{Introduction}
\label{intro}

In a remarkable paper from 1986 \ct{bekenstein} J. Bekenstein proposed the
intriguing idea about a gravity-assisted spontaneous symmetry breaking of
electro-weak (Higgs) type without invoking unnatural (according to
Bekenstein's opinion) ingredients like negative mass squared and a quartic
self-interaction for the Higgs field. By considering a model of gravity interacting
with a standard Klein-Gordon field (with small positive mass squared and
without self-interaction) coupled conformally to the scalar
curvature he managed to obtain a prototype of dynamically induced Higgs-like
spontaneous symmetry breaking scalar potential. A similar approach was further
worked out in Ref.~\refcite{moniz-etal}.

Motivated by Bekenstein's idea, in the present essay we will consider a
non-standard model of gravity coupled to a neutral scalar ``inflaton'' $\vp$ as
well as to a $U(1)$-charged $SU(2)$ iso-doublet scalar field $\s$ with a
standard positive mass squared and no self-interaction, as well as to 
$SU(2)\times U(1)$ gauge fields. The essential non-standard feature of this
model is employing non-Riemannian space-time volume forms -- alternative generally
covariant integration measure densities defined in terms of auxiliary
antisymmetric tensor gauge fields independent of the pertinent Riemannian
metric\foot{The method of non-Riemannian space-time volume forms,
originally proposed in Refs.~\refcite{TMT-orig-1,TMT-orig-2} has been employed in a broad
variety of gravity-matter, supergravity and strings/branes models to provide
plausible solutions for dynamical generation of cosmological constant, the
supersymmetric Higgs effect \ct{susyssb}, unified description of early 
universe inflation and present day dark energy \ct{emergent}, unified description
of dark energy and dark matter as different manifestations of a single entity 
\ct{dusty-1,dusty-2}}. 
Although being almost pure-gauge degrees of freedom (see the remark
in Section 2 below), the non-Riemannian space-time volume forms trigger a
series of important features unavailable in ordinary gravity-matter models
with the standard Riemannian volume-form (given by the square-root of the determinant
of the Riemannian metric): (i) The ``inflaton'' $\vp$ develops a remarkable
effective scalar potential in the Einstein frame possessing an infinitely
large flat region for large negative $\vp$ describing the ``early'' universe
evolution; (ii) In the absence of the $SU(2)\times U(1)$ iso-doublet scalar field, the
``inflaton'' effective potential has another infinitely large flat region for 
large positive $\vp$ describing the ``late'' post-inflationary 
(dark energy dominated) universe;
(iii) Inclusion of the $SU(2)\times U(1)$ iso-doublet scalar field $\s$ introduces a
drastic change in the total effective scalar potential in the post-inflationary 
universe -- the effective potential as a function of $\s$ acquires exactly
the electro-weak Higgs-type spontaneous symmetry breaking form.

\section{Gravity-Gauge-Field-Matter Model With Two Independent Non-Riemannian 
Volume-Forms}
\label{TMMT}
Our starting point is the following non-standard gravity-gauge-field-matter system 
with an action of the general form involving two independent non-Riemannian
volume-forms 
generalizing the model studied in Ref.~\refcite{emergent} and whose gauge-field-matter
part has an internal $SU(2)\times U(1)$ gauge symmetry (for simplicity 
we will use units where the Newton constant is taken as $G_{\rm Newton} = 1/16\pi$):
\be
S = \int d^4 x\,\P_1 (A) \Bigl\lb R - 2\L_0 \frac{\P_1 (A)}{\sqrt{-g}} 
+ L^{(1)} \Bigr\rb + \int d^4 x\,\P_2 (B) \Bigl\lb L^{(2)} + 
\frac{\P (H)}{\sqrt{-g}}\Bigr\rb \; .
\lab{TMMT}
\ee
In \rf{TMMT} and in the sequel the following notations are used:

\begin{itemize}
\item
$\P_{1}(A)$ and $\P_2 (B)$ are two independent non-Riemannian volume-forms, 
\textsl{i.e.}, alternative generally covariant integration measure densities on 
the underlying space-time manifold:
\be
\P_1 (A) = \frac{1}{3!}\vareps^{\m\n\k\l} \pa_\m A_{\n\k\l} \quad ,\quad
\P_2 (B) = \frac{1}{3!}\vareps^{\m\n\k\l} \pa_\m B_{\n\k\l} \; ,
\lab{Phi-1-2}
\ee
defined in terms of field-strengths of two auxiliary 3-index antisymmetric
tensor gauge fields. $\P_{1,2}$ take over the role of the standard Riemannian 
integration measure density 
$\sqrt{-g} \equiv \sqrt{-\det\Vert g_{\m\n}\Vert}$ in terms of the space-time
metric $g_{\m\n}$.
\item
$R = g^{\m\n} R_{\m\n}(\G)$ and $R_{\m\n}(\G)$ are the scalar curvature and the 
Ricci tensor in the first-order (Palatini) formalism, where the affine
connection $\G^\m_{\n\l}$ is \textsl{a priori} independent of the metric $g_{\m\n}$.
$\L_0$ is a small 
parameter later to be
identified with the present epoch small observable cosmological constant.
\item
$L^{(1)}$ is the sum of two scalar field Lagrangians:
\be
L^{(1)} = -\h g^{\m\n} \pa_\m \vp \pa_\n \vp - V_1(\vp) -
g^{\m\n} \bigl(\nabla_\m \s_a)^{*}\nabla_\n \s_a - V_0 (\s) \; ,
\lab{L-1}
\ee
where $\vp$ denotes the neutral ``inflaton'' inert under $SU(2)\times U(1)$
and $\s \equiv (\s_a)$ 
is a complex $SU(2)\times U(1)$ iso-doublet scalar field
with the isospinor index $a=+,0$ indicating the corresponding $U(1)$ charge.
The gauge-covariant derivative in \rf{L-1} acting on the iso-doublet scalar $\s$ reads:
\be
\nabla_\m \s = 
\Bigl(\pa_\m - \frac{i}{2} \t_A \cA_\m^A - \frac{i}{2} \cB_\m \Bigr)\s \; ,
\lab{cov-der}
\ee
with $\h \t_A$ ($\t_A$ -- Pauli matrices, $A=1,2,3$) indicating the $SU(2)$ 
generators and $\cA_\m^A$ ($A=1,2,3$)
and $\cB_\m$ denoting the corresponding $SU(2)$ and $U(1)$ gauge fields. 
The pertinent scalar field potentials are:
\be
V_1 (\vp) = f_1 \exp \{-\a\vp\} \quad ,\quad
V_0 (\s) = m^2_0\, \s^{*}_a \s_a \; ,
\lab{V-1-0}
\ee
where $\a, f_1$ are dimensionful positive parameters, and $V_0 (\s)$ is just the
standard mass term for the iso-doublet $\s_a$ with positive mass squared.
\item
$L^{(2)}$ is the sum of a second ``inflaton'' Lagrangian plus the canonical
Lagrangians for the $SU(2)$ and $U(1)$ gauge fields $\cA^A_\m$, $\cB_\m$:
\be
L^{(2)} = -\frac{b}{2} e^{-\a\vp} g^{\m\n} \pa_\m \vp \pa_\n \vp + U(\vp) 
- \frac{1}{4g^2} F^2(\cA) - \frac{1}{4g^{\pr\,2}} F^2(\cB) \; ,
\lab{L-2}
\ee
where:
\be
U(\vp) = f_2 \exp \{-2\a\vp\} \; ,
\lab{U}
\ee
with $f_2$ another dimensionful positive parameter, whereas $b$ is a
dimensionless one, and (all indices $A,B,C = (1,2,3)$):
\br
F^2(\cA) \equiv F^A_{\m\n} (\cA) F^A_{\k\l} (\cA) g^{\m\k} g^{\n\l} \quad ,\quad
F^2(\cB) \equiv F_{\m\n} (\cB) F_{\k\l} (\cB) g^{\m\k} g^{\n\l} \; ,
\lab{F2-def} \\
F^A_{\m\n} (\cA) = 
\pa_\m \cA^A_\n - \pa_\n \cA^A_\m + \eps^{ABC} \cA^B_\m \cA^C_\n \quad ,\quad
F_{\m\n} (\cB) = \pa_\m \cB_\n - \pa_\n \cB_\m \; .
\lab{F-def}
\er
\item
$\P (H)$ indicates the dual field strength of a third auxiliary 3-index antisymmetric
tensor gauge field:
\be
\P (H) = \frac{1}{3!}\vareps^{\m\n\k\l} \pa_\m H_{\n\k\l} \; ,
\lab{Phi-H}
\ee
whose presence is crucial for non-triviality of the model. 
\end{itemize}

\textbf{Remark.} The systematic canonical Hamiltonian treatment of \rf{TMMT} 
(see {\em Appendix}) and, more
generally, of any gravity-matter models built with non-Riemannian volume-forms
\ct{buggy} shows that the auxiliary 3-index tensor gauge fields $A_{\m\n\k}$,
$B_{\m\n\k}$, $H_{\m\n\k}$ are almost pure-gauge, \textsl{i.e.}, they do not
introduce any new propagating field degrees of freedom except for a few
free integration constants in their respective equations of motion (see
\rf{A-B-H-eqs}-\rf{integr-const} below).

Let us note that the requirement for invariance (with the exception of the mass 
term $V_0 (\s)$ of the iso-doublet $\s_a$) under the following global Weyl-scale 
symmetry:
\br
g_{\m\n} \to \l g_{\m\n} \;\; ,\;\; \vp \to \vp + \frac{1}{\a}\ln \l \;\;,\;\; 
A_{\m\n\k} \to \l A_{\m\n\k} \;\; ,\;\; B_{\m\n\k} \to \l^2 B_{\m\n\k} \; ,
\lab{scale-transf} \\
\G^\m_{\n\l} \;,\; H_{\m\n\k} \;,\; \s_a \; ,\; \cA^A_\m \; ,\; \cB_\m \;
- \; {\rm inert} \; , \phantom{aaaaaaaaaaaa}
\nonu
\er
uniquely fixes the structure of the non-Riemannian-measure gravity-gauge-field-matter
action \rf{TMMT}. In particular, for the same reason we have multiplied by an 
appropriate exponential factor the ``inflaton'' kinetic term in $L^{(2)}$ \rf{L-2}.

Following Ref.~\refcite{emergent} we will now derive the effective Einstein-frame form
of the dynamics described by \rf{TMMT}.

Variation of the action \rf{TMMT} w.r.t. affine connection $\G^\m_{\n\l}$:
\be
\int d^4\,x\,\sqrt{-g} g^{\m\n} \Bigl(\frac{\P_1}{\sqrt{-g}}\Bigr) 
\(\nabla_\k \d\G^\k_{\m\n} - \nabla_\m \d\G^\k_{\k\n}\) = 0 
\lab{var-G}
\ee
shows 
that $\G^\m_{\n\l}$ becomes a Levi-Civita connection
$ \G^\m_{\n\l} = \G^\m_{\n\l}({\bar g}) = 
\h {\bar g}^{\m\k}\(\pa_\n {\bar g}_{\l\k} + \pa_\l {\bar g}_{\n\k} 
- \pa_\k {\bar g}_{\n\l}\)$
w.r.t. to the Weyl-rescaled metric ${\bar g}_{\m\n}$:
\be
{\bar g}_{\m\n} = \chi_1 \, g_{\m\n} \quad ,\quad
\chi_1 \equiv \frac{\P_1 (A)}{\sqrt{-g}} \; .
\lab{bar-g}
\ee

Variation of \rf{TMMT} w.r.t. auxiliary tensor gauge fields
$A_{\m\n\l}$, $B_{\m\n\l}$ and $H_{\m\n\l}$ yields the equations (using the
short-hand notation $\chi_1$ from \rf{bar-g}):
\be
\pa_\m \Bigl\lb R + L^{(1)} - 4\L_0 \chi_1\Bigr\rb = 0 \quad, \quad
\pa_\m \Bigl\lb L^{(2)} + \frac{\P (H)}{\sqrt{-g}}\Bigr\rb = 0 
\quad, \quad \pa_\m \Bigl(\frac{\P_2 (B)}{\sqrt{-g}}\Bigr) = 0 \; ,
\lab{A-B-H-eqs}
\ee
whose solutions read:
\br
\frac{\P_2 (B)}{\sqrt{-g}} \equiv \chi_2 = {\rm const} \;\; ,\;\;
R + L^{(1)} - 4\L_0 \chi_1\ = - M_1 = {\rm const} \; , 
\nonu \\
L^{(2)} + \frac{\P (H)}{\sqrt{-g}} = - M_2  = {\rm const} \; .
\lab{integr-const}
\er
Here $M_1$ and $M_2$ are arbitrary dimensionful and $\chi_2$
arbitrary dimensionless integration constants. 

Now, varying \rf{TMMT} w.r.t. $g_{\m\n}$ and using relations \rf{integr-const} 
we obtain:
\be
\chi_1 \Bigl\lb R_{\m\n} + \h\( g_{\m\n}L^{(1)} - T^{(1)}_{\m\n}\) -
\L_0 \chi_1 g_{\m\n}\Bigr\rb -
\h \chi_2 \Bigl\lb T^{(2)}_{\m\n} + g_{\m\n} M_2 \Bigr\rb = 0 \; ,
\lab{pre-einstein-eqs}
\ee
where $\chi_1$ and $\chi_2$ are the same as in \rf{bar-g} and \rf{integr-const},
respectively, and $T^{(1,2)}_{\m\n}$ are the energy-momentum tensors of the
matter-gauge-field Lagrangians \rf{L-1}-\rf{L-2} with the standard definitions:
$T^{(1,2)}_{\m\n} = g_{\m\n} L^{(1,2)} - 2 \pa L^{(1,2)}/\pa g^{\m\n}$.


Taking the trace of Eqs.\rf{pre-einstein-eqs} and using again the second relation 
\rf{integr-const} we solve for the scale factor $\chi_1$ \rf{bar-g}:
\be
\chi_1 = 2 \chi_2 \frac{T^{(2)}/4 + M_2}{L^{(1)} - T^{(1)}/2 - M_1} 
\quad ,\quad T^{(1,2)} = g^{\m\n} T^{(1,2)}_{\m\n} \; .
\lab{chi-1}
\ee

Using the second relation \rf{integr-const}, the gravity equations \rf{pre-einstein-eqs} 
can be put in the Einstein-like form:
\be
R_{\m\n} - \h g_{\m\n}R = \h g_{\m\n}\(L^{(1)} + M_1 - 2 \L_0 \chi_1\)
+ \h\(T^{(1)}_{\m\n} - g_{\m\n}L^{(1)}\)
+ \frac{\chi_2}{2\chi_1} \Bigl\lb T^{(2)}_{\m\n} + g_{\m\n} M_2 \Bigr\rb \; ,
\lab{einstein-like-eqs}
\ee

Now, using expression \rf{chi-1} for $\chi_1$ and following 
the same steps as in Ref.~\refcite{emergent} we can bring Eqs.\rf{einstein-like-eqs} 
into the standard form of Einstein 
equations for the rescaled  metric ${\bar g}_{\m\n}$ \rf{bar-g}, 
\textsl{i.e.}, the Einstein-frame gravity equations: 
\be
R_{\m\n}({\bar g}) - \h {\bar g}_{\m\n} R({\bar g}) = \h T^{\rm eff}_{\m\n}
\quad ,\quad
T^{\rm eff}_{\m\n} = g_{\m\n} L_{\rm eff} - 2 \partder{}{g^{\m\n}} L_{\rm eff} 
\; ,
\lab{eff-einstein-eqs}
\ee
with effective energy-momentum tensor corresponding 
to the following effective Einstein-frame matter Lagrangian:


\be
L_{\rm eff} = A(\vp,\s) X + B(\vp) X^2 - U_{\rm eff}(\vp,\s) 
- {\bar g}^{\m\n} \bigl(\nabla_\m \s_a)^{*}\nabla_\n \s_a
- \frac{\chi_2}{4g^2} {\bar F}^2(\cA) - \frac{\chi_2}{4g^{\pr\,2}} {\bar F}^2(\cB)
\; .
\lab{L-eff-final}
\ee

In \rf{L-eff-final} the following notations are used:

\begin{itemize}
\item
$X$ is the standard short-hand notations for the ``inflaton'' kinetic term:
\be
X \equiv - \h {\bar g}^{\m\n}\pa_\m \vp \pa_\n \vp \; .
\lab{X-def}
\ee
\item
The coefficient functions in the ``k-essence''-type 
\ct{k-essence-1,k-essence-2,k-essence-3,k-essence-4} expression 
(first two terms on the r.h.s. of \rf{L-eff-final} with {\em non-linear} functional 
dependence on $X$ \rf{X-def}) 
are given by:
\br
A(\vp,\s) \equiv 1 + \h b e^{-\a\vp}\,
\frac{V_1 (\vp) + V_0 (\s) - M_1}{U(\vp) + M_2}
\nonu\\
= 1 + \h b e^{-\a\vp}\,
\frac{f_1 e^{-\a\vp} + m_0^2 \s_a^{*}\s_a - M_1}{f_2 e^{-2\a\vp} + M_2} \; ; 
\lab{A-def}
\er
and
\be
B(\vp) \equiv - \frac{1}{4}\,\frac{\chi_2 b^2 e^{-2\a\vp}}{U(\vp) + M_2}
= - \frac{1}{4}\,\frac{\chi_2 b^2 e^{-2\a\vp}}{f_2 e^{-2\a\vp} + M_2} \; .
\lab{B-def}
\ee
\item
The full effective scalar field potential reads:
\br
U_{\rm eff} (\vp,\s) \equiv 
\frac{\Bigl\lb V_1(\vp) + V_0 (\s) - M_1 \Bigr\rb^2}{
4\chi_2 \Bigl\lb U(\vp) + M_2\Bigr\rb} + 2\L_0
\nonu\\
= \frac{\(f_1 e^{-\a\vp} + m_0^2 \s_a^{*}\s_a - M_1\)^2}{4\chi_2\,\Bigl\lb 
f_2 e^{-2\a\vp} + M_2\Bigr\rb} + 2\L_0\; ,
\lab{U-eff}
\er
where in\rf{A-def}-\rf{U-eff} the explicit forms of $V_1 (\vp), V_0 (\s)$ and 
$U(\vp)$ \rf{L-1}-\rf{L-2} are inserted.
\item
In the last line of \rf{L-eff-final} the Einstein-frame versions of the
gauge-field Lagrangians are used, where:
\be
{\bar F}^2(\cA) \equiv F^A_{\m\n} (\cA) F^A_{\k\l} (\cA) {\bar g}^{\m\k} {\bar g}^{\n\l} 
\quad ,\quad
{\bar F}^2(\cB) \equiv F_{\m\n} (\cB) F_{\k\l} (\cB) {\bar g}^{\m\k} {\bar g}^{\n\l} \; .
\lab{bar-F2-def}
\ee
\end{itemize}

\section{Effective Scalar Potential -- Infinitely Large Flat Regions and
Emergent Higgs-like Behavoir}
\label{flat+higgs}

A remarkable feature of the effective Einstein-frame scalar potential \rf{U-eff}
is that it possesses an infinitely large flat region for large negative
$\vp$ and is independent of $\s$ there:
\br
U_{\rm eff}(\vp,\s) \simeq U_{(-)} \equiv \frac{f_1^2}{4\chi_2\, f_2} + 2\L_0 
\; , \phantom{aaaaaaaaaaaa}
\lab{minus-region} \\
A(\vp,\s) \simeq A_{(-)} \equiv 1+\h b \frac{f_1}{f_2} \;\; ,\;\;
B(\vp) \simeq B_{(-)} \equiv - \chi_2 \frac{b^2}{4f_2} \; .
\nonu
\er
Thus, for large negative values of the ``inflaton'' the effective matter
Lagrangian \rf{L-eff-final} reads:
\br
L_{\rm eff}^{(-)} = A_{(-)} X + B_{(-)} X^2 - U_{(-)} + L\lb\s,\cA,\cB\rb \; ,
\lab{L-minus} \\
L\lb\s,\cA,\cB\rb \equiv
- {\bar g}^{\m\n} \bigl(\nabla_\m \s_a)^{*}\nabla_\n \s_a
- \frac{\chi_2}{4g^2} {\bar F}^2(\cA) - \frac{\chi_2}{4g^{\pr\,2}} {\bar F}^2(\cB) \; ,
\lab{L-sigma-gauge-def}
\er
\textsl{i.e.}, the $SU(2)\times U(1)$ iso-doublet scalar $\s$ becomes massless.

For large positive values of $\vp$ we obtain
$A(\vp,\s) \simeq A_{(+)} = 1$ , $ B(\vp) \simeq B_{(+)} = 0$, and:
\be
U_{\rm eff}(\vp,\s) \simeq U_{(+)}(\s) \equiv 
\frac{1}{4\chi_2\, M_2} \Bigl( m_0^2 \s_a^{*} \s_a - M_1\Bigr)^2 + 2\L_0 \; ,
\lab{plus-region}
\ee
so that the effective matter Lagrangian \rf{L-eff-final} becomes 
(using notation \rf{L-sigma-gauge-def}):
\be
L_{\rm eff}^{(+)} = X  - U_{(+)}(\s) + L\lb\s,\cA,\cB\rb \; .
\lab{L-plus}
\ee

In the case of absent $SU(2)\times U(1)$ scalar and gauge fields, a case
already discussed in detail in Ref.~\refcite{emergent}, the purely ``inflaton'' effective
potential $U_{\rm eff}(\vp) = \( f_1 e^{-\a\vp} - M_1\)^2
\llb 4\chi_2\,\bigl( f_2 e^{-2\a\vp} + M_2\bigr)\rrb^{-1}$
possesses another infinitely large flat region for large positive values of $\vp$: 
$U_{\rm eff}(\vp) \simeq U_{(+)} \equiv M_1^2\, \(4\chi_2\, M_2\)^{-1}$
and the shape of $U_{\rm eff}(\vp)$ 
is depicted in Fig.1.

\begin{figure}
\begin{center}
\includegraphics{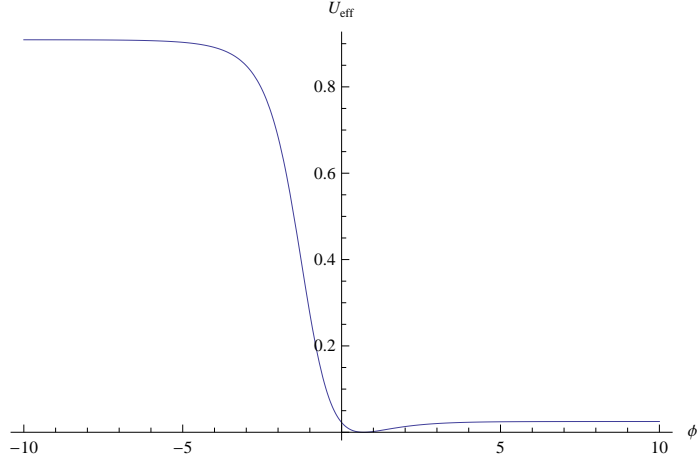}
\caption{Qualitative shape of the effective ``inflaton'' potential $U_{\rm eff}(\vp)$
as function of $\vp$, $M_1 > 0$.}
\end{center}
\end{figure}

It was found in Ref.~\refcite{emergent} that this effective ``inflaton'' potential
with two infinitely large flat regions with vastly different scales upon
appropriate choice for the values of the parameters accomodates the
description simultaneously of the ``early'' universe evolution (the inflationary
epoch) corresponding to the first flat region for large negative
$\vp$-values (the left one on Fig.1), as well as the ``late'' post-inflationary 
(dark energy dominated) universe corresponding to the second flat region for large 
positive $\vp$-values (the right one on Fig.1).

The presence of the $SU(2)\times U(1)$ scalar and gauge fields will not affect 
significantly the inflationary dynamics in the ``early'' universe (when $\vp$ runs 
on the left higher flat region of $U_{\rm eff}(\vp)$, Fig.1), and where
the $SU(2)\times U(1)$ iso-doublet $\s$-field is massless, see 
\rf{L-sigma-gauge-def})) --
this dynamics is governed by the purely kinetic ``k-essence'' ``inflaton'' part 
$A_{(-)} X + B_{(-)} X^2 - U_{(-)}$ of \rf{L-minus}.

However, in the post-inflationary epoch (when $\vp$ runs on the right lower flat 
region of $U_{\rm eff} (\vp)$, Fig.1) the presence of the $SU(2)\times U(1)$
iso-doublet $\s$ contributes very significantly to the full effective scalar potential
\rf{plus-region}, which emerges in an exactly Higgs-like form:
\br
U_{(+)}(\s) = 
\frac{1}{4\chi_2\, M_2} \Bigl( m_0^2 \s_a^{*} \s_a - M_1\Bigr)^2 + 2\L_0
\equiv
\l\, \bigl(\s_a^{*}\s_a\bigr)^2 - \m^2 \s_a^{*}\s_a + {\rm const} \; ,
\lab{higgs-like} \\
\l = \frac{m_0^4}{4\chi_2\,M_2} \quad ,\quad 
\m^2 = \frac{M_1\,m_0^2}{2\chi_2\,M_2} \; . \phantom{aaaaaaaaaaaaaaaa}
\lab{higgs-param}
\er
Spontaneous $SU(2)\times U(1)$ symmetry breakdown occurs at the vacuum value:
\be
|\s_{\rm vac}| = \frac{1}{m_0}\sqrt{M_1} \; , 
\lab{s-vac}
\ee
with a small residual cosmological constant $\L_0$, which is identified with the 
current epoch observable cosmological constant. Let us also note that
in the post-inflationary epoch according to \rf{L-plus} and \rf{higgs-like}
the massless ``inflaton'' $\vp$ (entering the kinetic term $X$ \rf{X-def} only) 
looses all interactions with the $SU(2)\times U(1)$ iso-doublet $\s$
(which has now become the Higgs field), thus avoiding undesirable fifth-force problems.

The dependence of the Higgs-like parameters \rf{higgs-param} and the Higgs-like 
v.e.v. \rf{s-vac} on the integration constants $M_{1,2}$ and $\chi_2$ explicitly
reveals the nature of the Higgs-like spontaneous gauge symmetry breaking 
\rf{higgs-like} as a 
gravity-assisted 
dynamically generated phenomenon in the post-inflationary epoch. 
It is triggered exclusively through the special non-Riemannian volume-form dynamics
in the original gravity-gauge-field-matter action \rf{TMMT} leading to the
remarkable two-flat-region shape of the ``inflaton'' effective potential (Fig.1).
Let us stress again that in the ``early'' universe there is no spontaneous
breaking of $SU(2)\times U(1)$ gauge symmetry and the Higgs-like iso-doublet
scalar field $\s$ is massless there \rf{L-minus}-\rf{L-sigma-gauge-def}.

We conclude with a note on the plausible numerical values for some of the
parameters involved. In Ref.~\refcite{emergent} it was argued that it is natural
to associate $M_1 \sim M^4_{EW}$, $M_2 \sim M^4_{Pl}$ and $\chi_2 \sim 10^{-1}$,
where $M_{EW}$ and $M_{Pl}$ are the electroweak and Planck scales, respectively.
Then, from \rf{s-vac} we see that it is natural also to associate the bare
mass of the $SU(2)\times U(1)$ iso-doublet $\s$-field $m_0 \sim M_{EW}$, so
that we will have for the Higgs-like v.e.v. $|\s_{\rm vac}| \sim M_{EW}$  
conforming to the standard electroweak phenomenology.

\section*{Acknowledgements} 
We gratefully acknowledge support of our collaboration through the academic exchange 
agreement between the Ben-Gurion University and the Bulgarian Academy of Sciences.
S.P. and E.N. have received partial support from European COST actions
MP-1210 and MP-1405, respectively, as well from Bulgarian National Science
Fund Grant DFNI-T02/6. We also thank the referee for useful remarks and typo
corrections.
\appendix

\section{Canonical Hamiltonian Treatment of Gravity-Matter Theories with
Non-Riemannian Volume-Forms}

Let us briefly discuss the application of the canonical Hamiltonian formalism 
to the gravity-matter model based on two non-Riemannian spacetime volume-forms 
\rf{TMMT}. For convenience we introduce the following short-hand notations for the
field-strengths \rf{Phi-1-2}, \rf{Phi-H} of the auxiliary 3-index antisymmetric gauge 
fields (the dot indicating time-derivative): 
\br
\P_1 (A) = \Adot + \pa_i A^i \quad, \quad 
A = \frac{1}{3!} \vareps^{ijk} A_{ijk} \;\; ,\;\;
A^i = - \h \vareps^{ijk} A_{0jk} \; ,
\lab{A-can} \\
\P_2 (B) = \Bdot + \pa_i B^i \quad, \quad 
B = \frac{1}{3!} \vareps^{ijk} B_{ijk} \;\; ,\;\;
B^i = - \h \vareps^{ijk} B_{0jk} \; ,
\lab{B-can} \\
\P (H) = \Hdot + \pa_i H^i \quad, \quad 
H = \frac{1}{3!} \vareps^{ijk} H_{ijk} \;\; ,\;\;
H^i = - \h \vareps^{ijk} H_{0jk} \; ,
\lab{H-can}
\er

The canonical momenta conjugated to \rf{A-can}-\rf{H-can} read:
\br
\pi_A = R + L^{(1)} - \frac{4\L_0}{\sqrt{-g}} \(\Adot + \pa_i A^i\) \; ,
\lab{can-momenta-A} \\
\pi_B = L^{(2)} + \frac{1}{\sqrt{-g}}\(\Hdot + \pa_i H^i\) \;\; ,\;\;
\pi_H = \frac{1}{\sqrt{-g}}\(\Bdot + \pa_i B^i\) \; ,
\lab{can-momenta-B-H}
\er
and:
\be
\pi_{A^i} = 0 \quad,\quad \pi_{B^i} = 0 \quad,\quad \pi_{H^i} = 0 \; .
\lab{can-momenta-zero}
\ee
The latter imply that $A^i, B^i, H^i$ are in fact Lagrange multipliers
for certain first-class Hamiltonian constraints 
(see Eqs.\rf{Ham-can}-\rf{1st-class-constr} below). 

Using the short-hand notation $(u,\udot)$ to collectively denote 
the set of the basic gravity-matter canonical variables 
$(u)=\bigl(g_{\m\n}, \vp, \s, \cA^A_\m ,\cB_m \bigr)$ and their respective 
velocities, we have for their respective canonically conjugated momenta:
\be
p_u = \(\Adot + \pa_i A^i\) \partder{}{\udot}\Bigl( R + L^{(1)}\Bigr) +
\(\Bdot + \pa_i B^i\)\partder{}{\udot} L^{(2)}
\lab{can-momenta-u}
\ee

Now, using relations \rf{can-momenta-A}-\rf{can-momenta-u} we arrive at the
following Dirac-constrained canonical Hamiltonian of the model \rf{TMMT}:
\br
\cH_{\rm can} = p_u \udot  
- \frac{\sqrt{-g}}{8\L_0} \Bigl( R + L^{(1)} -\pi_A\Bigr)^2 - \pi_H \sqrt{-g} L^{(2)}
\nonu \\
+ \sqrt{-g} \pi_H\,\pi_B - \pa_i A^i \pi_A - \pa_i B^i \pi_B - \pa_i H^i \pi_H \; .
\lab{Ham-can}
\er
The last three terms in \rf{Ham-can} show that, indeed, the auxiliary 
gauge field components $A^i, B^i, H^i$ (with vanishing canonically conjugated
momenta \rf{can-momenta-zero}) are Lagrange multipliers for the Dirac first class 
constraints:
\be
\pi_A = - M_1 = {\rm const} \;\; ,\;\; \pi_B = - M_2 = {\rm const} 
\;\; ,\;\; \pi_H = \chi_2 = {\rm const} \;; ,
\lab{1st-class-constr}
\ee
which are the canonical Hamiltonian counterparts of Lagrangian constraint
equations of motion \rf{integr-const}.

To conclude, the canonical Hamiltonian treatment of \rf{TMMT} reveals the 
(almost) pure-gauge nature of the auxiliary 3-index antisymmetric tensor gauge
fields $A_{\m\n\l},\, B_{\m\n\l},\, H_{\m\n\l}$ -- building blocks of
the non-Riemannian spacetime volume-form formulation of the modified gravity-matter
model \rf{TMMT}. Namely, the canonical momenta $\pi_A,\, \pi_B,\, \pi_H$ 
conjugated to the ``magnetic'' parts $A,B,H$ \rf{A-can}-\rf{H-can}
of the auxiliary 3-index antisymmetric tensor gauge fields are constrained
through Dirac first-class constraints \rf{1st-class-constr}
to be constants identified with the arbitrary 
integration constants $\chi_2,\, M_1,\, M_2$ \rf{integr-const} arising within the 
Lagrangian formulation of the model. The canonical momenta 
$\pi_A^i,\, \pi_B^i,\, \pi_H^i$ conjugated to the ``electric'' parts $A^i,B^i,H^i$ 
\rf{A-can}-\rf{H-can} of the auxiliary 3-index antisymmetric tensor gauge field
are vanishing \rf{can-momenta-zero}, which makes the latter canonical Lagrange 
multipliers for the above Dirac first-class constraints.

Thus, the method of employing non-Riemannian spacetime volume-forms in
gravity-matter theories does not introduce any additional propagating
field-theoretic degrees of freedom apart from the standard ones.


\end{document}